\documentclass[aps,twocolumn,pra,superscript,floatfix,showpacs,footinbib]{revtex4-1}
\usepackage{amssymb}

\usepackage[pdftex]{graphicx}
\usepackage{dcolumn}
\usepackage{bm}
\usepackage{amsmath}
\usepackage{array}
\usepackage{color}
\usepackage{float}
\usepackage{subfigure}
\usepackage{dsfont}
\usepackage{txfonts}
\usepackage{wasysym}
\usepackage{multirow}
\usepackage{sidecap}
\usepackage{xcolor,cancel}
\usepackage[normalem]{ulem}

\newcommand{\dn}{\downarrow}

\newcommand{\up}{\uparrow}

\newcommand{\ket}[1]{{|{#1}\rangle}}
\newcommand{\bra}[1]{{\langle{#1}|}}

\usepackage{hyperref}
\hypersetup{
    colorlinks,%
    citecolor=blue,%
    linkcolor=blue,%
    urlcolor=blue
}

\begin{document} 


\title{Robustness of Stark many-body localization in the $J_1$-$J_2$ Heisenberg model}

\author{E. Vernek}
\affiliation{Instituto de F\'isica, Universidade Federal de 
Uberl\^andia, Uberl\^andia, Minas Gerais 38400-902, Brazil.}


\date{\today}

\begin{abstract}
Stark many-body localization (SMBL) is a phenomenon observed in interacting systems with a nearly uniform spatial gradient applied field. Contrasting to the traditional many-body localization phenomenon, SMBL does not require disorder. Here we investigate SMBL in a spin-$1/2$ described by a Heisenberg model including a next-nearest-neighbor exchange coupling. By employing an exact diagonalization approach  and time evolution calculation we analyze both level spacing ratio (LSR) statistics of the Hamiltonian model as well as the dynamics of the system from a given initial state. Our results reveals that for zero field in our finite system, LSR statistics suggest localization while the dynamics shows thermalization, which has been attributed to a finite-size effect. Slightly nonuniform field gradient, LSR statistic predictions agree very well with the dynamics of the physical quantities indicating delocalization and localization for small and large field gradient, respectively. More interestingly, we find that localization is robust in the presence of  next-nearest-neighbor coupling in the Hamiltonian. Moreover, this coupling can be tuned to enhance SMBL in the system, meaning that  localized regimes can be obtained for smaller field gradient as compared to the traditional nearest-neighbor isotropic Heisenberg model.

\end{abstract}
\maketitle
\section{Introduction}
The ubiquitous phenomenon of localization in quantum systems has attracted a great deal of attention since its prediction in  the seminal paper by Anderson~\cite{PhysRev.109.1492,doi:10.1063/1.3206091,Kramer_1993}. Originally predicted in disordered non interacting quantum systems, the phenomenon has become even more attracting in interacting systems where disorder is also capable of localizing many-body states ~\cite{annurev-conmatphys-031214-014726,annurev-conmatphys-031214-014701,10.1002/andp.201700169,Smith2016}. In these systems, the phenomenon has gained a special terminology, namely many-body quantum localization (MBL).
While initially  investigated in quantum systems, the phenomenon can be as well observed in classical systems~\cite{doi:10.1080/13642818508240619} including propagating electromagnetic waves~\cite{10.1038/nphys1101,10.1038/nphoton.2013.30}.  

One remarkable feature of MBL in closed quantum systems is the lack of thermalization.  If a closed system, fully described by a Hamiltonian  $H$, is initially (at $t=0$) in some initial state $\mid \phi(t=0)\rangle$, the time evolution of that state is given by $\mid \phi(t)\rangle={\cal U}(t)\mid \phi(t=0)\rangle$, where ${\cal U}(t)=e^{-iHt/\hbar}$ is the unity time evolution operator. This unitary time evolution preserves all physical quantities associated with any observable $\hat O $ commuting with $H$. Suppose that in the initial state the expectation value of the operator $\hat O$ defined locally such that $\hat O=\sum_jO_j$ (where $j$ corresponds to positions in the system) has expectation value zero, i.e., $\langle \hat O\rangle_\phi=\langle\phi(t=0)\mid\hat{O}\mid \phi(t=0)\rangle=0$.  We can imagine, in addition, that in the initial state the local expectation value $\langle\phi(t=0)\mid\hat{O}_j\mid \phi(t=0)\rangle=-1$ for all $j$ in one half of the system and $\langle\phi(t=0)\mid\hat{O}_j\mid \phi(t=0)\rangle=1$ for $j$ in the other half. Such an initial state is highly inhomogeneous, therefore in such a state the system is out of thermodynamical equilibrium. We say that the system thermalizes if at long time scales, upon time evolution, the expectation value of the operator $\hat O_j$ is uniformly distributed across the system, which in the present example would vanish everywhere. 

In disordered closed interacting systems, despite the unitary character of time evolution, the phenomenon of MBL prevents thermalization. In the localized regime, the system is said to be \emph{non-ergodic}, while the thermal regime is called \emph{ergodic}. In the nonergodic regime,  the system keeps memory of the initial state under time evolution, while  in the ergodic regime the information about the initial state is somehow  erased during the evolution. Quite  intuitively, there exists a transition from the ergodic to nonergodic regime when disorder increases progressively from zero~\cite{PhysRevB.82.174411,PhysRevB.103.174203,PhysRevB.104.174201}. Over the past few years important progress has been made on observation of MBL~\cite{Schreiber842,Choi1547,PhysRevX.7.041047,PhysRevResearch.3.033043}. However, given the intricate character of localization in quantum many-body systems, this phenomenon is still under intense investigation~\cite{PhysRevResearch.3.013023,PhysRevLett.127.030601,LOZANONEGRO2021111175}. 
  
Recently, lack of thermalization in quantum systems has been predicted even in the absence of disorder~\cite{PhysRevLett.122.040606,van_Nieuwenburg9269,doi:10.1098/rsta.2011.0363}. The main ingredient in this case is the presence of potentials with nearly uniform spatial gradients. The resulting features predicted in the localized phase of such models are very much similar to those of MBL systems, which includes includes level spacing ratio (LSR) statistics and memory of initial states~\cite{PhysRevLett.122.040606,van_Nieuwenburg9269,PhysRevLett.124.110603,PhysRevLett.123.030602}. This phenomenon was suggestively called Stark many-body localization (SMBL)~\cite{PhysRevLett.122.040606}, and was later predicted to be observed in fermion~\cite{PhysRevLett.126.210601,PhysRevA.103.023323,PhysRevLett.123.030602}, spin ~\cite{PhysRevB.103.L100202} as well as in topological systems~\cite{PhysRevResearch.2.023067}. These theoretical predictions have  been recently confirmed by experimental signatures of SMBL  in  trapped-ions quantum simulators~\cite{morong2021observation}.
More recently, SMBL has been investigated in a one-dimensional Heisenberg model with a uniform spatial field gradient along the chain~\cite{PhysRevB.103.L100202}. It was shown that the system dos not thermalize if initialized in certain classes of initial states containing domain walls separating opposite magnetization regions along the chain~\cite{PhysRevB.103.L100202}. It was shown that in the strongly localized regime the domain walls (if at low density) do not melt as time evolves, a remarkable signature of quantum localization. The lack of thermalization of certain initial states in these system has been associated to the  concept \emph{fragmentation} of Hilbert  space~\cite{PhysRevX.10.011047,PhysRevB.101.174204,PhysRevB.102.054206,PhysRevB.103.134207,PhysRevB.104.155117}. Under this concept, the time evolution of this type of states avoids some regions of the Hilbert space, thus preventing the system from thermalization. As such, this phenomenon violates the established eigenstate thermalization hypothesis (ETH) in its \emph{strong sense}, meaning that thermalization  thermalization depends on the particular initial state. Bearing this in mind, if for a given system Hamiltonian  thermalization occurs only for some initial states, the system may not thermalize if initiated in the same state but evolving under a slightly  modified Hamiltonian. In this work we investigate SMBL in a spin-1/2 Heisenberg whose exchange coupling is extended to next-nearest neighbors. This model is known in the literature  as the $J_1$-$J_2$ model or Majumdar-Ghosh Hamiltonian and describes rich quantum frustrated phases~\cite{OKAMOTO1992433,Bursill_1995}. These phases are controlled solely by the ratio $J_1/J_2$. While it is known that the ground state is very sensitive to change of theses couplings, the question  of whether SMBL is affected by details fine tuning of these parameters is hitherto unknown. 

Within this context, the central question we want to address is how the introduction of nearest neighbor interaction affects SMBL in the system induced by a nearly uniform gradient magnetic field. We show that for a  uniform magnetic field gradient, LSR distribution deviates from the one  expected for Poisson statistics even though the initial state does not thermalize, which has been previously associated to many exact degeneracies in the spectrum \cite{PhysRevLett.122.040606}. Interestingly, for strictly zero field, {LSR statistics follows  the Poisson statistics while the  ergodicity is expected~\cite{PhysRevB.91.081103,PhysRevB.98.174202}. We  attributed this to a strong finite-size effect in the LSR analysis}.  For a small slope in the field gradient, however, standard LSR distribution statistics are obtained. In this case, for zero magnetic field   the distribution follows  the  expected Wigner-Dyson surmise  while in the  localized regime it exhibits the  distribution~\cite{PhysRevB.93.041424} expected for Poisson processes. Further, we follow Polyakov {\it et al.} \cite{PhysRevB.103.L100202} and focus on an initial state containing a single island of positive magnetization and monitor if the system keeps a memory of the initial state upon time evolution governed by the Hamiltonian with different values of couplings. By applying a time-dependent variation principle method~\cite{PhysRevB.102.094315} we confirm the localized and  delocalized regimes of the system predicted in the literature for zero magnetic field. This shows that application of LRS analysis for a uniform field gradient is not straightforward~\cite{PhysRevLett.122.040606}. Moreover, we show that SMBL is very robust against changes in the couplings of the Hamiltonian. More interestingly, our results show that there are regions in the $J_1$-$J_2$ parameter space in which localization is enhanced as compared to the standard isotropic Heisenberg model.

This work is organized as follows: In Sec.~\ref{model} we introduce our model and methods, in Sec.~\ref{results} we present our numerical results and discussions. Finally, our concluding remarks are presented in  Sec.~\ref{conclusions}.

\section{Model and methods}\label{model}
For the sake of concreteness, in this work we consider a one-dimensional spin-1/2 chain described by the known $J_1$-$J_2$ Hamiltonian subject to a nonuniform magnetic field which can be written as
\begin{eqnarray}\label{H_J1J2}
H=J_1\sum_{j=1}^{N-1} {\bf S}_j \cdot {\bf S}_{j+1}+ J_2\sum_{j=1}^{N-2} {\bf S}_j \cdot {\bf S}_{j+2}+\sum_{j=1}^N h_jS_j^{z}.
\end{eqnarray}
Here ${\bf S}_i$ represent spin-1/2 operators, $J_1$ ($J_2$) represents the nearest (next-nearest) neighbor Heisenberg-like couplings and $h_i$ represents a magnetic field along the $z$ direction that varies along the chain. In particular, here we consider a magnetic field given by 
\begin{eqnarray}
h_j=h_0j+\frac{\gamma j^2}{N^2},
\end{eqnarray}
where $h_0$ is the gradient of the field and  $\gamma$ introduces a small nonlinearity in the field gradient.  Following Burssil {\it et al.} \cite{Bursill_1995} we parametrize the coupling as $(J_2,J_1)=(J_0\cos\theta,J_0\sin\theta )$ for $0< \theta < 2\pi$ which is sketched in Fig.~\ref{model}(a). In the case of  $h_0=\gamma=0$ the ground state of the Hamiltonian \eqref{H_J1J2} has been thoroughly studied over the past decades. Here, we are interested in thermalization processes which (in principle) involve  all eigenstates of the system. Since for any finite field gradient translation symmetry is absent in the system, here we set open boundary condition for all calculations. 
\begin{figure}[!t]
	\centering
	\subfigure{\includegraphics[clip,width=3.45in]{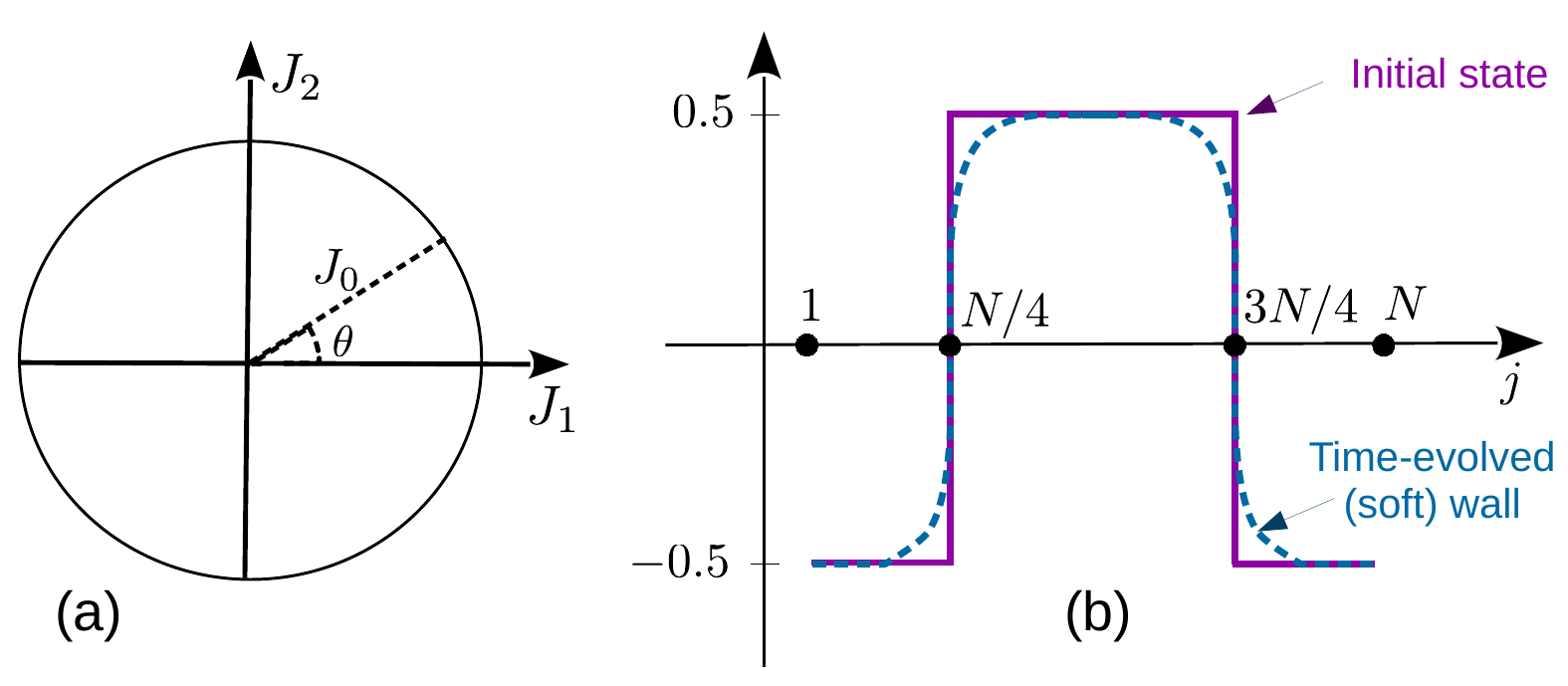}}
	\caption{(a) Graphical representation of the Hamiltonian couplings $J_1$ and $J_2$ parametrized in term of $J_0$ and $\theta$. (b) Sketch of an initial state with sharp domain walls (solid purple) and soft (blue dashed) walls after time evolution.} 
	\label{model}
\end{figure} 

\subsection{Time-evolution analysis and level spacing ratio statistics}
\label{Sec1A}

Our numerical analysis of thermalization is performed by time evolving two physical quantities, namely magnetization imbalance and entanglement entropy, that reveal thermalization of the system evolving from a given initial state $\mid \Psi_0 \rangle$. We will restrict ourselves to a initial state consisting of all spins \emph{down} but with an island of spins \emph{up} symmetrically placed in the middle of the chain. Such a state is schematically represented in Fig.~\ref{model}(b) by the solid purple line. More formally, for a chain of $N=8$ spin, such a state can be written as $\Psi_0=\mid \dn\dn \up \up \up \up \dn \dn \rangle$. This state possesses  zero total magnetization, $\sum_j \langle \Psi_0\mid S^z_j\mid \Psi_0\rangle=0$, but with the magnetic moment distributed nonuniformly along the chain, a characteristic of a nonequilibrium state. This type of state is one of those considered in Ref.~\cite{PhysRevB.103.L100202}. Since $\mid\Psi_0 \rangle $ is not an eigenstate of the system, it will time evolve nontrivially and in the thermalized situation the magnetization will be zero everywhere, resulting in $\bra{\Psi(t)}S_j^z\ket{\Psi(t)}\approx 0$ for all $j \in [1,N]$, where $\ket{\Psi(t)}$ is the state of the system evolved from $\ket{\Psi_0}$. The thermalization process will produce a softening of the domain wall as sketched in Fig.~\ref{model}(b) by the dashed  blue line.  

The time evolution of a quantum system  governed by a many-body Hamiltonian as our Eq.~\eqref{H_J1J2} is in general a very burdensome task. Here we adopt a time-dependent variation principle (TDVP) \cite{PhysRevB.94.165116} designed to the matrix product state (MPS) formalism~\cite{PAECKEL2019167998,SCHOLLWOCK201196}. One advantage of this method over the Suzuki-Trotter approach is the capability of dealing with Hamiltonian having non-local interactions, which is the case here for $J_2$ finite (see Ref.~\cite{PhysRevB.94.165116} for further discussion). To perform our calculation we use the iTensor library \cite{SciPostPhys.2.1.003}. 

To monitor the evolution of the system towards thermalization, we calculate the imbalance defined as
\begin{eqnarray}
{\cal I}(t)=\frac{4}{N}\sum_{j=1}^N \langle S_j^z(t)\rangle\langle S_j^z(0)\rangle,
\end{eqnarray} 
where $\langle S_i^z(t)\rangle=\bra{\Psi(t)}S_i^z\ket{\Psi(t)}$, calculated within the Schrödinger representation. The imbalance reveals whether the system keeps or loses information about local magnetization throughout the chain as the system evolves in time. In the strong localized regime of the system ${\cal I}(t)$ remains close to unity, while in the thermal regime it vanishes at long time scales. 

A second quantity used to examine how information on an initial state spreads over the system is the bipartite entanglement entropy~\cite{RevModPhys.82.277},
\begin{eqnarray}
S_\ell(t)=-\frac{1}{\ln 2}{\rm Tr}\left[\rho^{\ell}_A(t) \ln\rho^{\ell}_A(t)\right],
\end{eqnarray}
where $\rho^{\ell}_A={\rm Tr}_{\rm B}\rho^{\ell}_B$. Here, $A$ and $B$ are the subsystems obtained by  dividing the system at the left ($\ell=L$) or right ($\ell=R$) domain wall of the magnetization of the initial state, $\ket{\Psi_0}$.

Another approach commonly employed in the study of many-body localized systems is the energy LSR statistics. Within this approach, we define the LSR as $r_n=\min(\delta_n/\delta_{n+1},\delta_{n+1}/\delta_n)$, where $\delta_n=E_{n+1}-E_n$, in which $E_n$ is the $n$th eigenvalue of the Hamiltonian. Here, we obtain the  exact full spectrum of the Hamiltonian by exact diagonalization using Quspin library~\cite{SciPostPhys.2.1.003,10.21468/SciPostPhys.7.2.020}. By restricting ourselves to the Hilbert subspace of zero magnetization ($\langle S^z_{\rm tot}\rangle=0$), we are able to find the spectrum for a chain of length up to $L=16$ with relatively low computation effort. To obtain the distribution of LSR, we sample the total $r_n$ in $N{c}$ frequency class intervals and count how many $r_n$ lie within each interval. We then analyze the probability density  $P(r_n)$ as a function of $r_n$, which is usually taken as signature of the ergodicity of the system. Here $P(r_n)$ is such that $\sum_n P(r_n)/N_c=1$.

Finally, having the spectrum of the Hamiltonian at hand, for completeness, we calculate the density of states of the system defined usually as
\begin{eqnarray}\label{DOS}
{\rm DOS}(E)=\frac{1}{\cal N}\sum_k\delta(E-E_k),
\end{eqnarray}
in which ${\cal N}$ is the dimension of the Hilbert subspace corresponding to zero total magnetization.
For practical numerical calculation and to obtain a smooth curve, we represent the delta function as Lorentzian of width $\eta$. Taking $\eta=0.05$ is good enough to provide a good visualization of out  numerical results.

\section{Numerical results}\label{results}
To show our numerical result, let us take $J_0=1$ as our energy unity and set $\hbar=1$ so we can take $1/J_0$ as our time unity. Since we want to examine the system under linearly varying magnetic field along the chain alone and with an additional field varying quadratically, we will show these two cases side by side.

\subsection{Level spacing ratio statistics}

\begin{figure*}[!htbp]
	\centering
	\subfigure{\includegraphics[clip,width=3.5in]{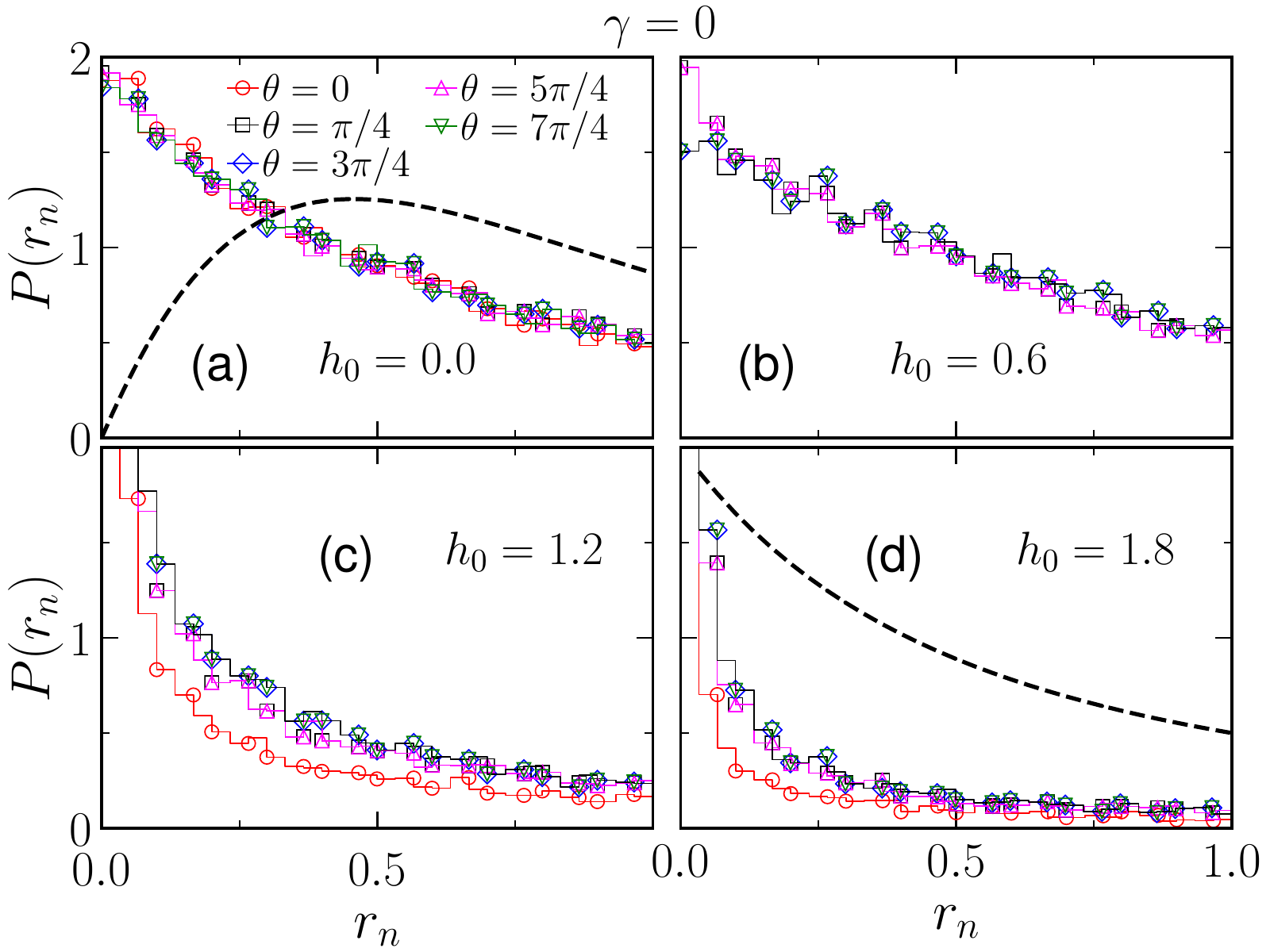} \includegraphics[clip,width=3.5in]{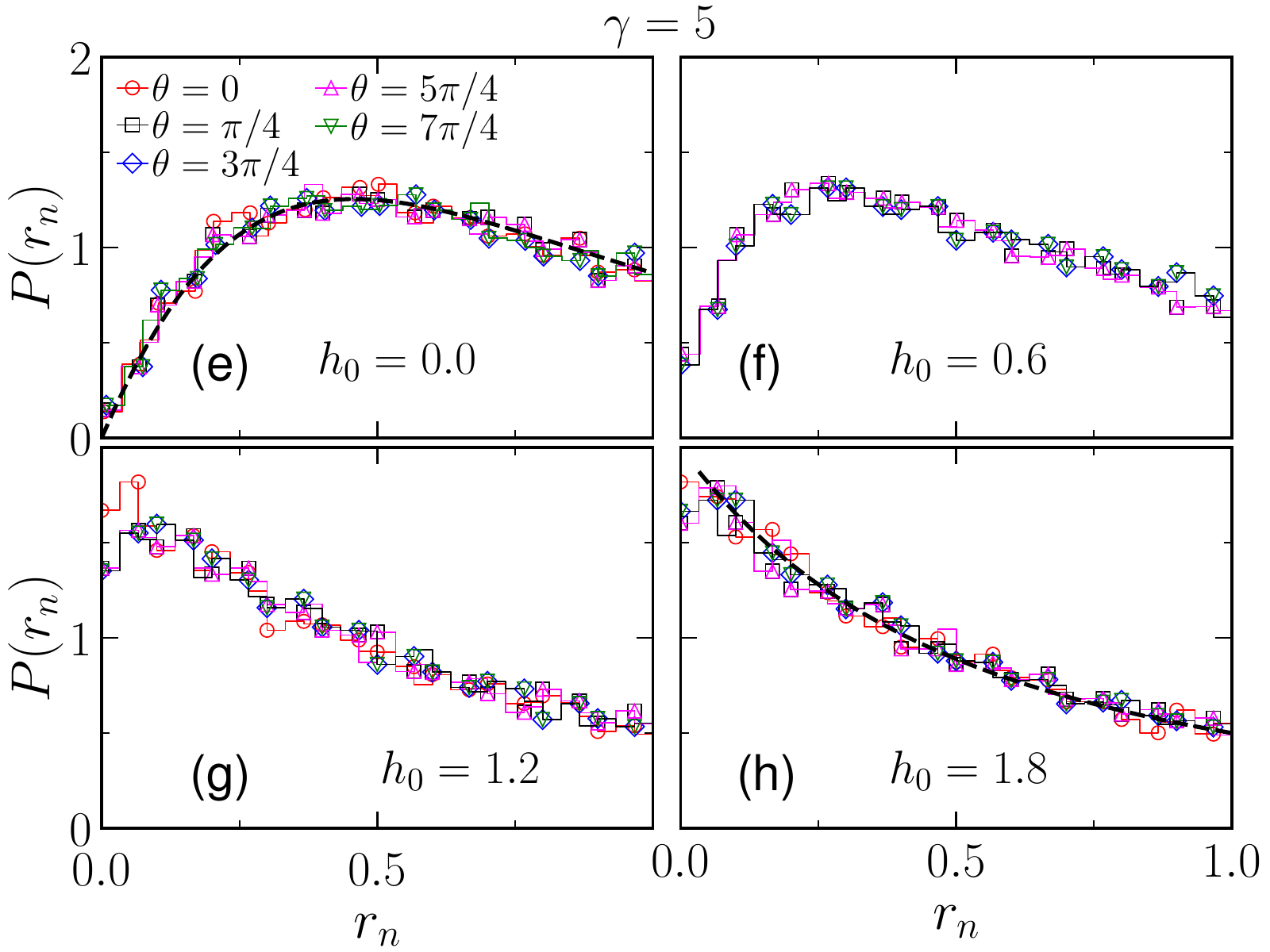}} 
	\caption{Level spacing ratio distribution, $\langle P(r_n)\rangle $ vs  $r_n$  for various values of $\theta$ for $h_0=0$ (a), $h_0=0.6$ (b), $h_0=1.2$ (c),  and $h_0=1.8$ (d) using $\gamma=0$ and the same from (e)--(h) for $\gamma=5$. For all panels we have used  $N=16$ and asymmetric case with $\gamma=5.0$. Dashed lines in panels (a) and (d)  and  (e) and (h) correspond to the distributions for Poisson statistics and the Wigner-Dyson, respectively.} 
	\label{fig1}
\end{figure*} 

Let us first take a look at the static physical quantities introduced above. Let us start by analyzing the probability distribution for the LSR $r_n$. In Fig.~\ref{fig1} we show $P(r_n)$ for various values field gradient $h_0$ and angle $\theta$. Figures.~\ref{fig1}(a)--\ref{fig1}(d) exhibit results for $\gamma=0$ while Figs.~\ref{fig1}(e)--\ref{fig1}(h) show the case of $\gamma=5$. Dashed lines in panels \ref{fig1}(a)--\ref{fig1}(d) correspond to the expected probability distribution $r_n$ for Poisson statistics, $P_{P}(r)=2/(1+r)^2$, while the ones shown in Figs.~\ref{fig1}(e) and \ref{fig1}(h) show the Wigner-Dyson distribution function,  $P_{WD}(r)=27(r+r^2)/4(1+r+r^2)^{5/2}$. These expressions have been derived in the context of random matrix theory~\cite{PhysRevLett.110.084101} and have later shown to fit quite nicely the LSR distributions of delocalized  and localized regimes of spinless fermion models~\cite{PhysRevLett.122.040606}\footnote{ {A pedagogical discussion on eigenvalues distribution of random matrices is pleasantly introduced in chapter 2 of Ref.~\cite{Livan_2018} }}.
 
Observing the behavior of $P(r_n)$ vs $r_n$ for the pure linear field ($\gamma=0$) [\ref{fig1}(a)--\ref{fig1}(d)] we first note that for $h_0=0$ [Figs.~\ref{fig1}(a)] the distribution follows quite nicely what is expected for the Poisson statistics. Nevertheless,  the initial state thermalizes, as we will confirm latter with the dynamics. These apparently conflicting results come from a strong finite size-effect in the LRS. Employing periodic boundary condition is not straightforward here as the system lacks translation symmetry  for any $h_0$ or $\gamma$ finite. The limit of very small $h_0$ and $\gamma$ is rather very peculiar. Furthermore, as pointed out in Ref.~\cite{PhysRevLett.122.040606}, the behavior of LRS analysis of SMBL in very long systems is still uncertain. 
As $h_0$ increases $P(r_n)$ becomes more suppressed for larger $r_n$, but with great contributions from small $r_n$, indicating large number of degenerate (or quasi degenerate) states. Many of these degeneracies are lifted for finite $\gamma$. Under the idea of fragmentation of the Hilbert space, we could imagine applying LRS within each fragment and recover the traditional LRS for the delocalized regime. However, as far as we know, such an approach is not yet attainable. Figure \ref{fig1}(e) and \ref{fig1}(f) show $P(r_n)$ vs $r_n$ for $\gamma=5.0$. We now observe that $P(r_n)$ follows nicely the Wigner-Dyson (WD) distribution for $h_0=0$ [Fig.~\ref{fig1}(e)] and evolves towards  Poisson statistics for finite $h_0$. Indeed, for $h_0=1.8$ we observe a nice collapse of all points on the dashed line, predicted to be obtained for the delocalized case. We have checked results for $h_0$ as larger as $h_0=4$ and noticed that this scenario remains unchanged. It is interesting to observe that these features are almost identical for all angles $\theta$, meaning that it is independent of the microscopic details of the Hamiltonian. 

\begin{figure}[!htbp]
	\centering
	\subfigure{\includegraphics[clip,width=3.25in]{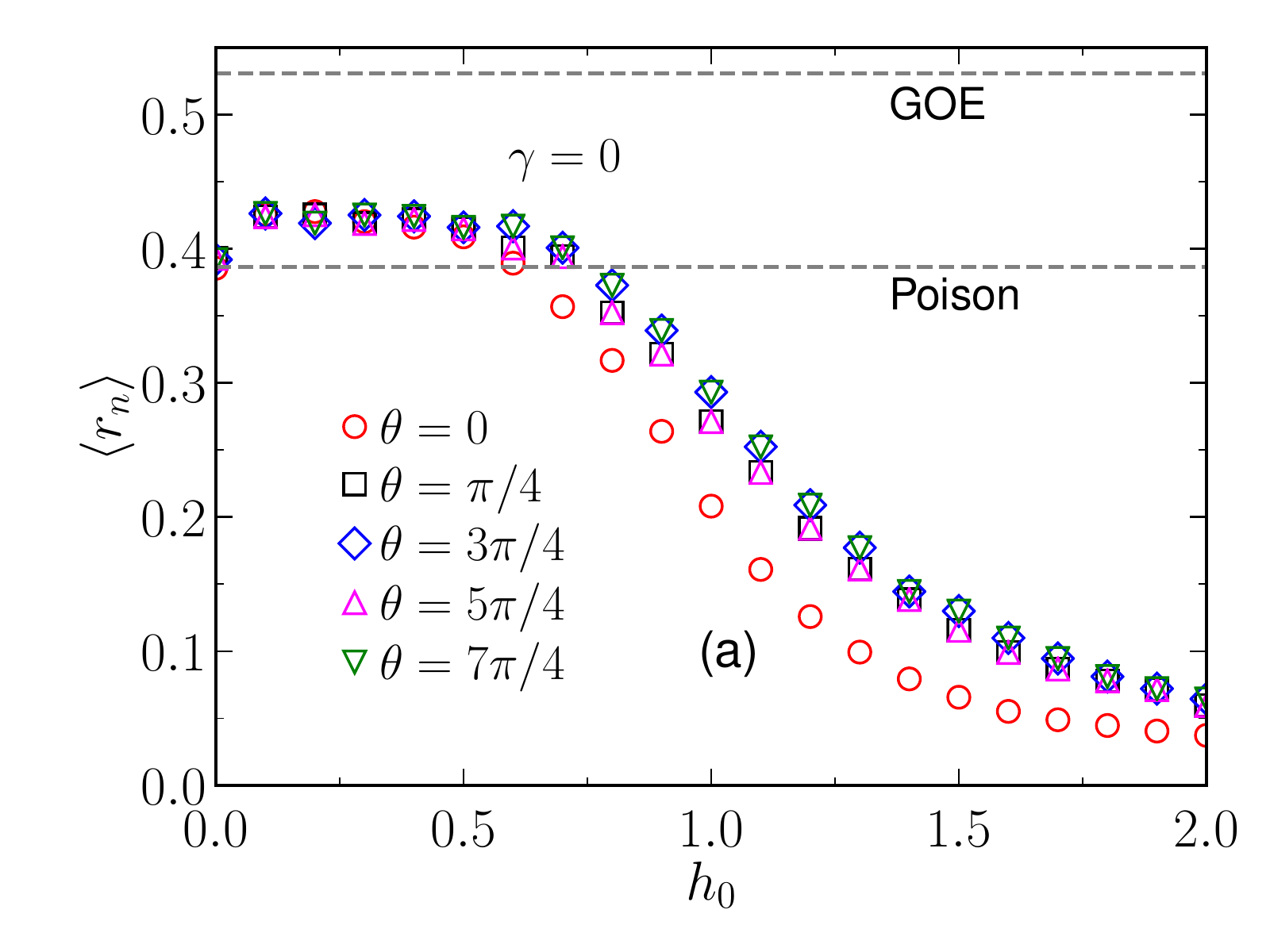}}
\vskip-0.48cm
	\subfigure{\includegraphics[clip,width=3.25in]{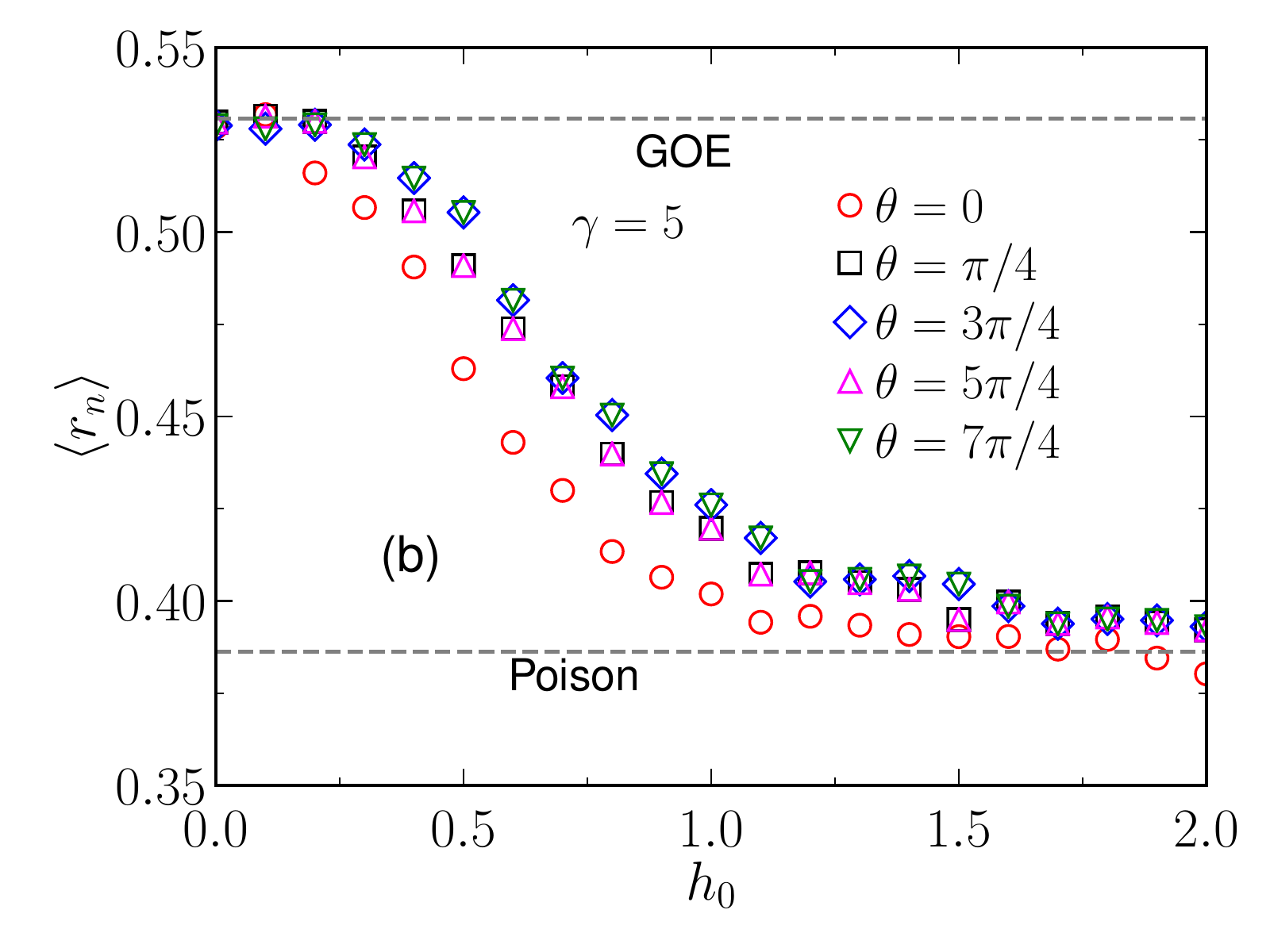}}       
	\caption{Mean value of level spacing LSR ratio $\bar r_n$ vs $h$ for various values of $\theta$ for $N=16$ and $\gamma=0$ (a) and $\gamma=5$ (b). Lower and upper dashed lines in each panel correspond the values expected for Poisson statistics and Gaussian orthogonal ensemble, respectively.} 
	\label{fig2}
\end{figure} 

In Fig.~\ref{fig2} we show the mean value of LSR, $\bar r_n$, for both $\gamma=0$  [Fig.~\ref{fig2}(a)] and $\gamma=5$ [Fig.~\ref{fig2}(b)]. We observe that the high density of $r_n$ for small $r_n$ results in $\bar r_n$ above the value expected for Poisson distribution ($\bar r_n \approx 0.38$). However, when $h_0$ increases $\bar r_n$ decreases rapidly because the distribution of $P(r_n)$ is peaked at small $r_n$. At finite $\gamma$, on the other hand, $\bar r_n$ starts close to the GOE  line ($\bar r_n \approx 0.53$ expected for the delocalized regime) and decreases as $h_0$ increases, setting down at $\bar r_n\approx 0.38$, also expected for the localized regime. Again, we observe similar behavior for all values of $\theta$. We note that there is an invariance in the results if we shift $\theta \rightarrow \theta+\pi$ observed for all results shown in  Figs.~\ref{fig1} and \ref{fig2}. This is trivially expected in the absence of magnetic field, since $H(\theta+\pi)\rightarrow -H(\theta)$ rendering $E_n(\theta+\pi)=-E_n(\theta)$. In the presence of field gradient, even though the spectrum is not symmetric, the spacing between adjacent levels still preserves the symmetry. Discussions on level statistics of this model in different spin sectors of the Hilbert have been discussed in Ref.~\cite{Poilblanc_1993}.
 
\begin{figure*}[!htbp]
	\centering
	\subfigure{\includegraphics[clip,width=3.5in]{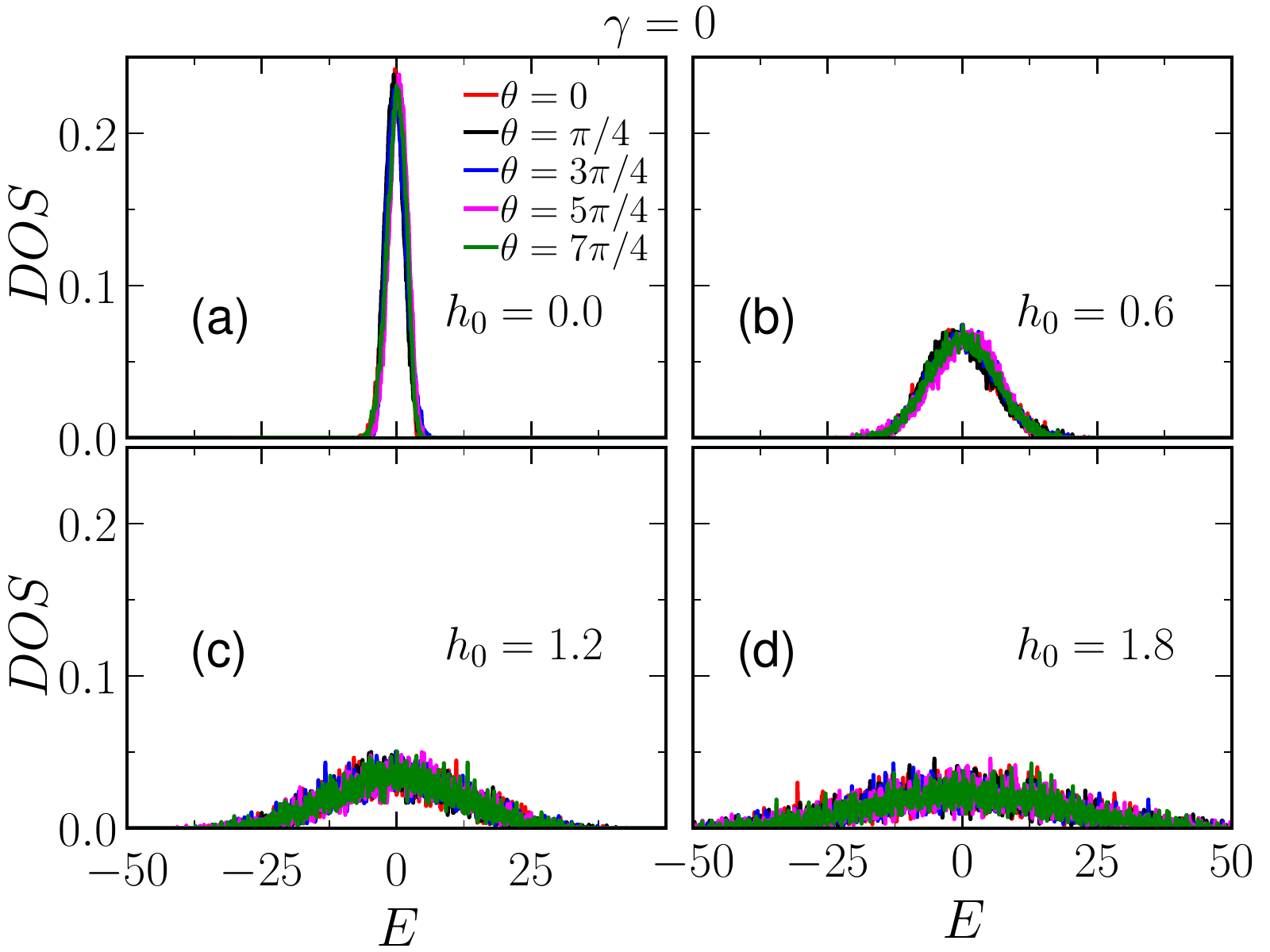}
    \includegraphics[clip,width=3.5in]{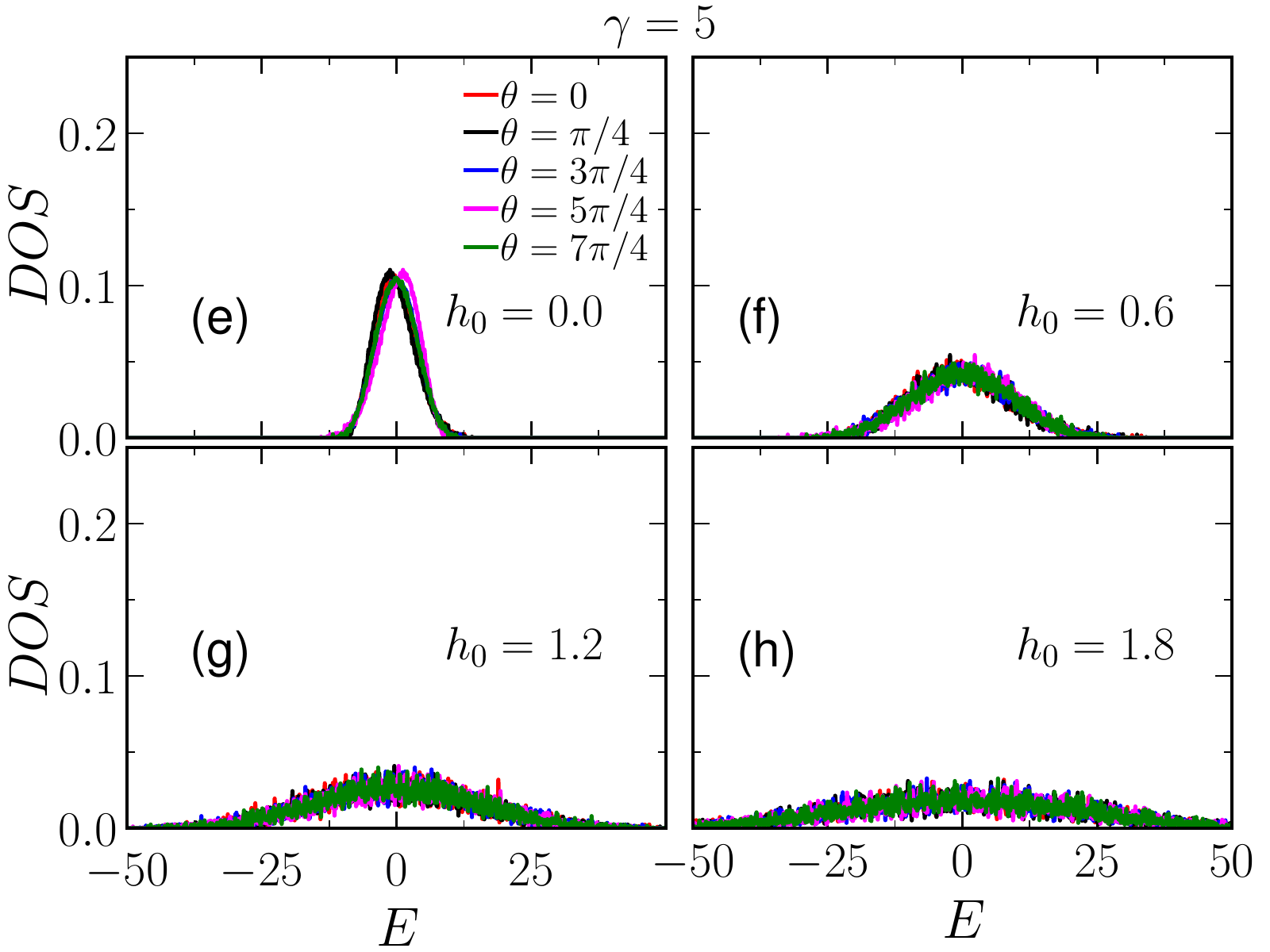}}     
	\caption{Density of states as a function of energy for various values of $\theta$ and for $h_0=0$ (a), $h_0=0.6$ (b), $h_0=1.2$ (c)  and $h_0=1.8$ (d) using $\gamma=0$. Panels (e)--(h) show the same as in (a)--(d) but for $\gamma=5$. For all panels we have used  $N=16$.} 
	\label{fig3}
\end{figure*} 

Before showing the results for the dynamics, in Fig.~\ref{fig3} we show the total density of states for the cases studied above. Like we did previously, Figs.~\ref{fig3}(a)--\ref{fig3}(d) show the DOS for $\gamma=0$, and Figs.~\ref{fig3}(e)--\ref{fig3}(h) for $\gamma=5$. We observe in both cases that the effect of the field is essentially to spread the energy levels to a wider range. The progressive suppression of the height of these curves guarantees their normalization.  Apart from a flatter shape of the curves observed for $\gamma=5$ as contrasted to those for $\gamma=0$, there is no striking differences between these two cases. With a careful look at the peaks of the DOS we see that they shift towards opposite directions as we change $\theta =\theta_0$  to $\theta=\theta_0+\pi$, rendering the symmetry in $r_n$ discussed above. 


\subsection{Dynamics}

\begin{figure*}[!htbp]
	\centering
	\subfigure{\includegraphics[clip,width=3.5in]{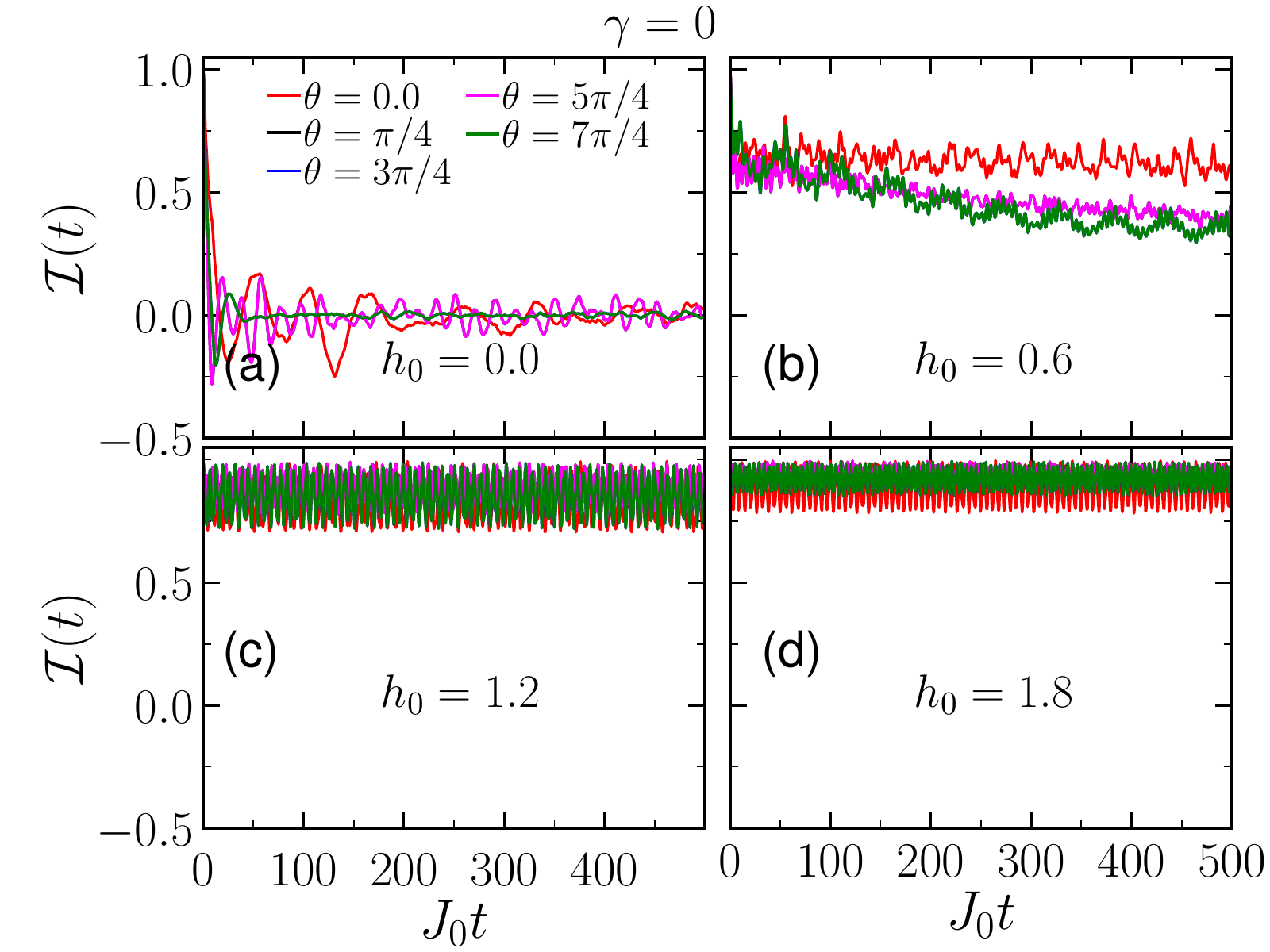} \includegraphics[clip,width=3.5in]{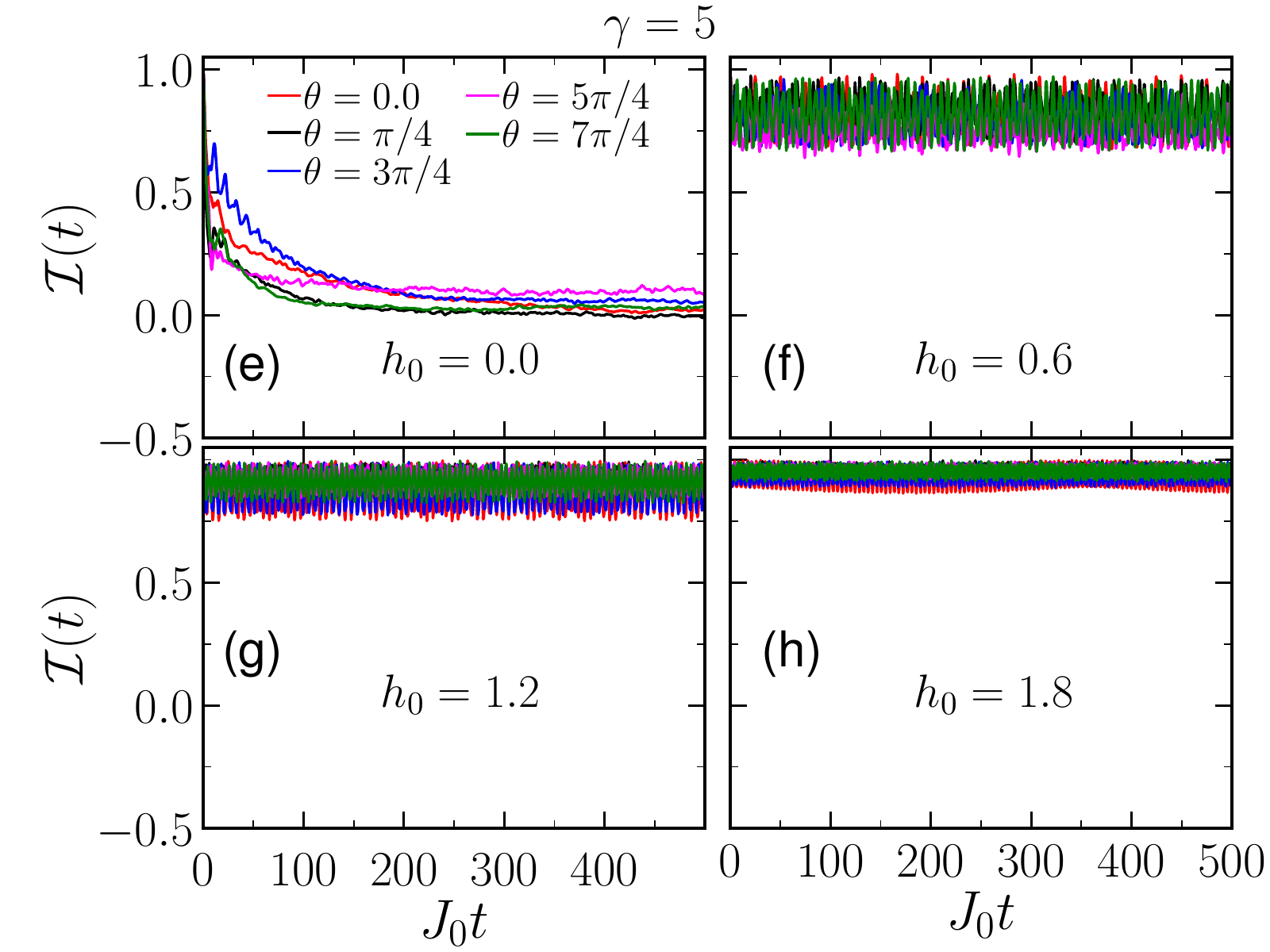}} 
	\caption{Imbalance, ${\cal I}(t)$ vs $t$ for various values of $\theta$ and for $h_0=0$ (a), $h_0=0.6$ (b), $h_0=1.2$ (c),  and $h_0=1.8$ (d) using $\gamma=0$. Panels (e)--(h) show the imbalance for the same values of parameters as in (a)--(d) but for $\gamma=5$. For all panels we have used  $N=16$. } 
	\label{fig4}
\end{figure*} 

To show the dynamics of relevant physical quantities discussed in the previous section, we set the initial state $\ket{\Psi_0}$  as defined in Sec.~\ref{Sec1A} and depicted in Fig.~\ref{model}(b). We first study the case of a chain of $L=16$, the same size of the one studied above. For all calculations in what follows we have truncated the TDVP algorithm to a maximum bond dimension $\chi=256$ (see discussion in Ref.~\cite{PhysRevB.102.094315}). An important limitation of this method is that in the delocalized regime the results may be sensitive to a choice of small $\chi$, as the bond dimension increases very quickly in along time evolution. It has been shown, however, that this problem is not severe even in the delocalized regime not very far from the crossover region \cite{PhysRevB.103.L100202,PhysRevB.98.174202}. In Fig.~\ref{fig4} we show the imbalance ${\cal I}(t)$ as a function of time up to $t=500/J_0$. We first note in Fig.~\ref{fig4}(a) that for $h_0=0$ ${\cal I}(t)$ decreases very quickly to zero, exhibiting small oscillations around zero. This confirms the thermalization of the system as it ``loses" information of the initial state, in the sense that it is spread across the many-body eigenstates involved in the time evolution. As $h_0$ increases to $0.6$ [Fig.~\ref{fig4}(b)] the imbalance decreases much slower. Interestingly, we note that ${\cal I}(t)$ decreases a little bit faster for $\theta$ finite (note that the red line is seemingly saturated since earlier times). We also note that the imbalances for $\theta$ and $\theta+\pi$ are identical. This symmetry were already noticed in the static quantities discussed above. Further increase of $h_0$ to $1.2$  and $1.8$ [Figs.~\ref{fig4}(c) and \ref{fig4}(d), respectively] show that ${\cal I}(t)$ remains close to unity, confirming a regime of quantum localization.
In Fig.~\ref{fig4}(e)--\ref{fig4}(h) we show the imbalance for the same set of parameters as before but with $\gamma=5$. We note that for $h_0=0$, ${\cal I}(t)$ evolves to almost zero at longer times. Which shows that, like the case of $\gamma=0$, the system loses information of the initial state. For larger values of $h_0$, shown in Figs.~\ref{fig4}(f)--\ref{fig4}(h), note that ${\cal I}(t)$ remains close to the unity for a long time. As compared to the case of $\gamma=0$ (and the same value of $h_0$) the long-time values of ${\cal I}(t)$ are much closer to 1, as compared to the case of $\gamma=0$. This suggests that the system is more localized for finite $\gamma$.

\begin{figure*}[!htbp]
	\centering
	\subfigure{\includegraphics[clip,width=3.5in]{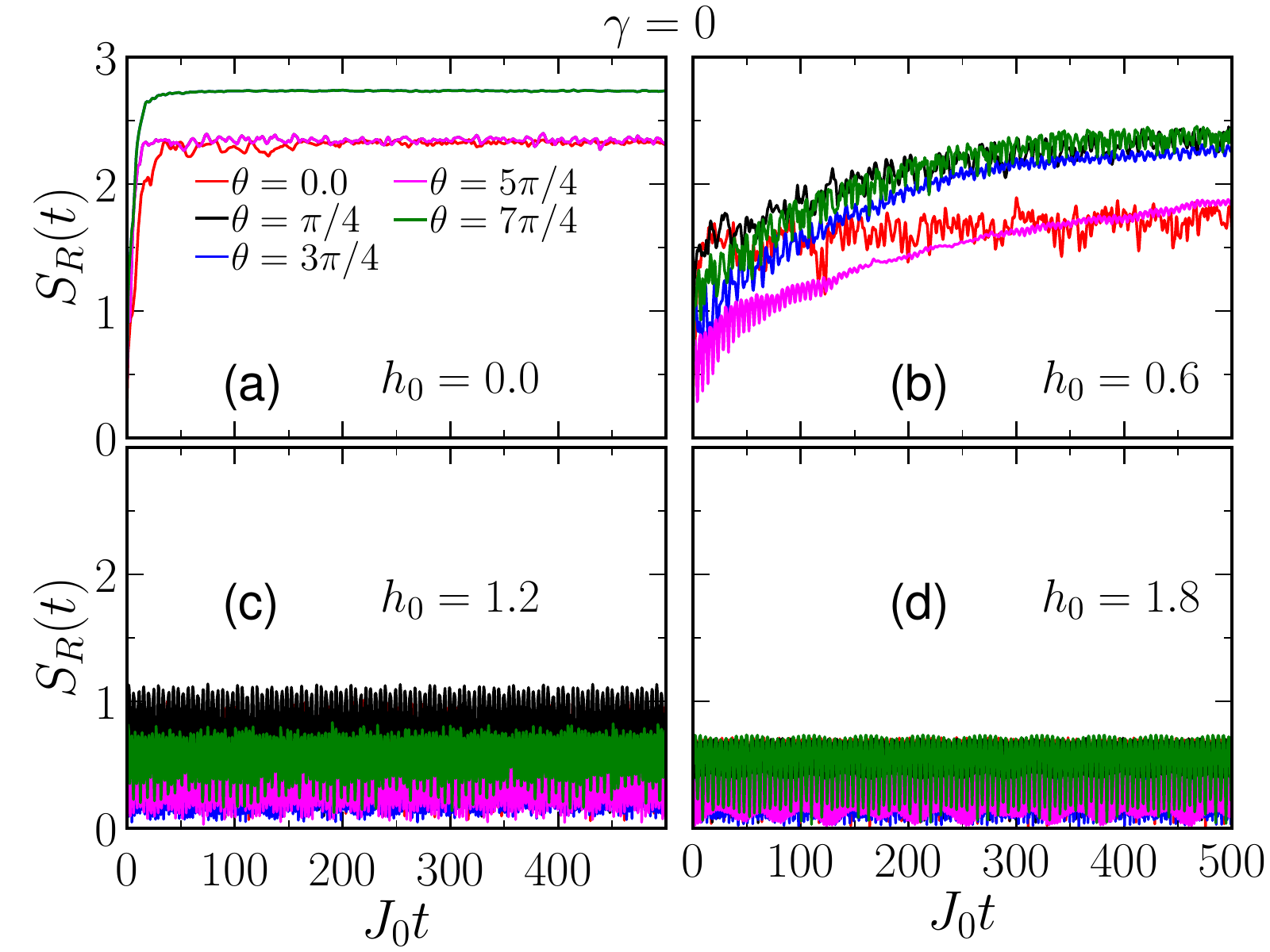} \includegraphics[clip,width=3.5in]{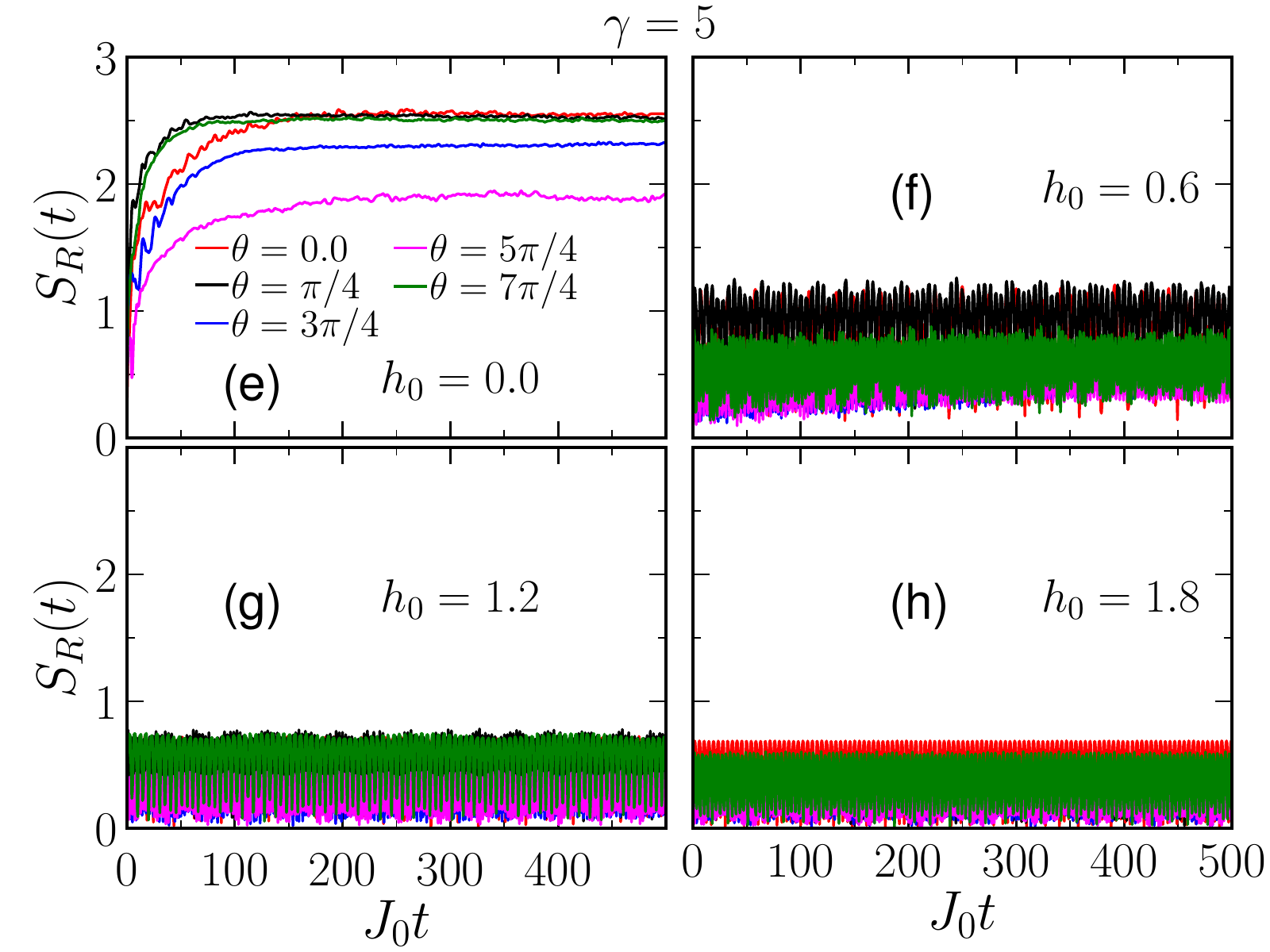}} 
	\caption{Entanglement entropy $S_R(t)$ vs $t$ calculated at the right domain wall  for various values of $\theta$ and for $h_0=0$ (a), $h_0=0.6$ (b), $h_0=1.2$ (c),  and $h_0=1.8$ (d) using $\gamma=0$. Panels (e)--(h) show the same  as in (a)--(d) but for $\gamma=5$. For all panels we have used  $N=16$. } 
	\label{fig5}
\end{figure*} 

To further analyze the physics o SMBL in the system, in Fig.~\ref{fig5} we show the entanglement entropy vs time calculated at position $j=3L/4$, on the right domain wall of the chain. We again note that $S_R(t)$  increases very quickly for $h_0=0$ [Fig.~\ref{fig5}(a)], saturating at a value $S(t \rightarrow \infty)$ above 2. For $h_0=0.6$ [Fig.~\ref{fig5}(b)] we observe that  $S(t)$ increases slower but we still see it saturating at a value about $S(t \rightarrow  \infty)$ around 2. It is interesting to see, however, that the entropy saturates much faster for $\theta=0$ as compared to finite $\theta$. This, again suggests that for finite $\theta$, the system tends to be less localized at least for small $h_0$. For larger values of  $h_0$ shown in Figs.~\ref{fig5}(c) and ~\ref{fig5}(d) $S(t)$ oscillates very rapidly around very small values, indicating that the information is not spread throughout the system, a characteristic of the localized regime. The reader may wonder why the  symmetry $S_R(\theta) =S_R(\theta+\pi)$ is not observed. It turns out that at large $t$, for $\theta =0$ the domain wall at $j=3L/4$ (on the right side) melts faster than the one at $j=L/4$ (the one on the left side). This situation is reversed for $\theta=\pi$ and swings back to the original shape as $\theta$ goes to $2\pi$. Therefore, since the entanglement entropy is defined locally, by the reasoning above, we conclude that  $S_R(\theta+\pi)=S_L(\theta)$. Our numerical calculations (not shown) indeed confirm this symmetry. Obviously, by changing the $h_0\rightarrow -h_0$ everything is reversed. 
The tendency of melting one domain wall faster than the other has been observed in Ref.~\cite{PhysRevB.103.L100202} and was ascribed to cooperation (or competition) between the magnetic field and the longitudinal component of the Heisenberg  coupling.

Figures ~\ref{fig5}(e)--\ref{fig5}(h) show the entropy for $\gamma=5$.  In this case, for $h_0=0$ we see that the entropy stabilizes to a value smaller than those observed for $\gamma=0$. This is because the small non uniform field somehow prevents full thermalization of the system. Moreover, we observe that the entropy is more sensitive to changes in the values of $\theta$ as compared to the case of $\gamma=0$. As $h_0$ increases we observe that thermalization is obtained for a smaller field. Note that for $h_0=0.6$ the entropy curves are already much lower than those obtained for the same value of $h_0$ for $\gamma=0$ shown in Fig.~\ref{fig5}(a)--\ref{fig5}(d). This confirms that a small gradient in the field favors SMBL in the system.

\begin{figure*}[!htbp]
	\centering
	\subfigure{\includegraphics[clip,width=3.5in]{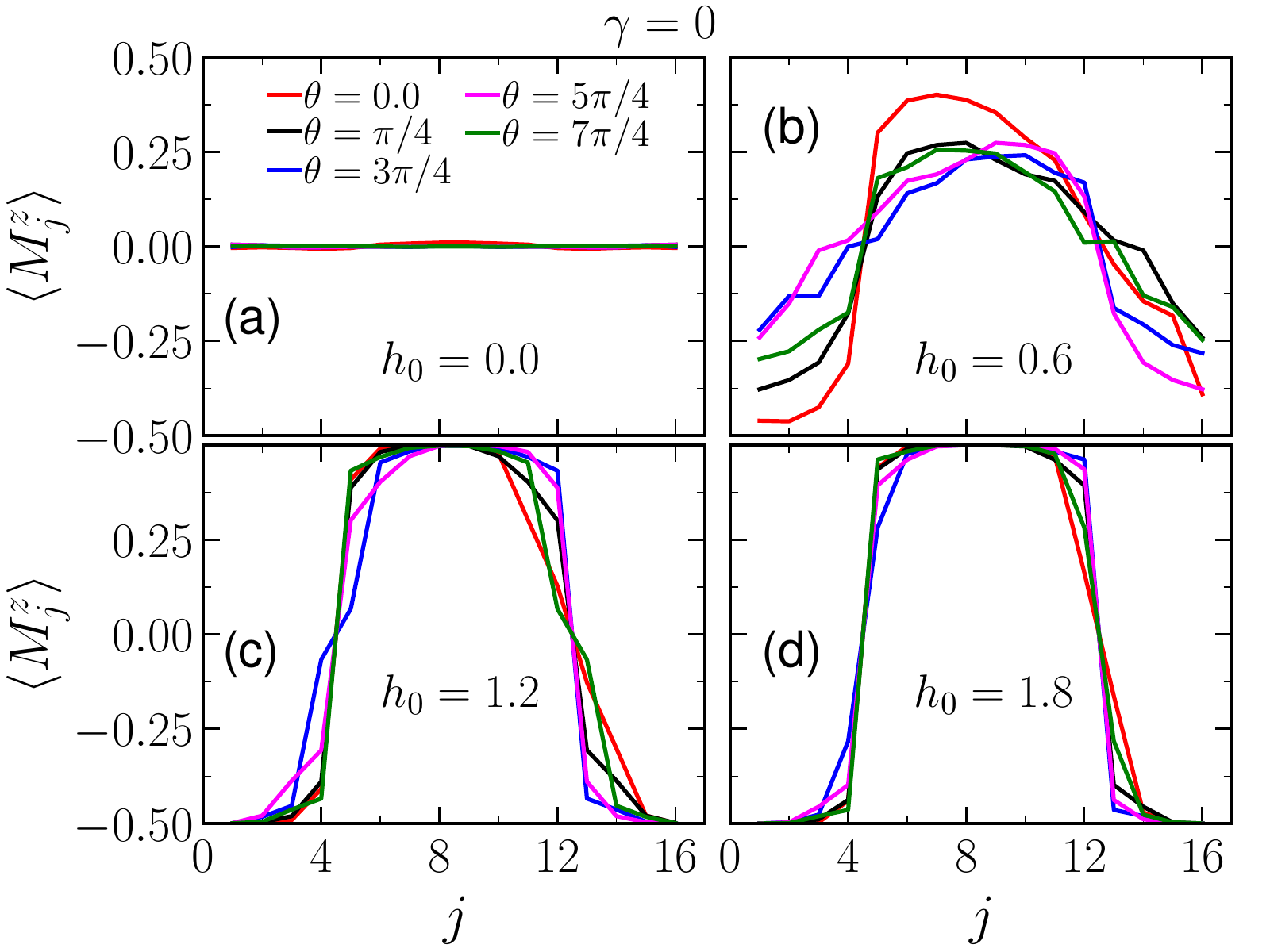} \includegraphics[clip,width=3.5in]{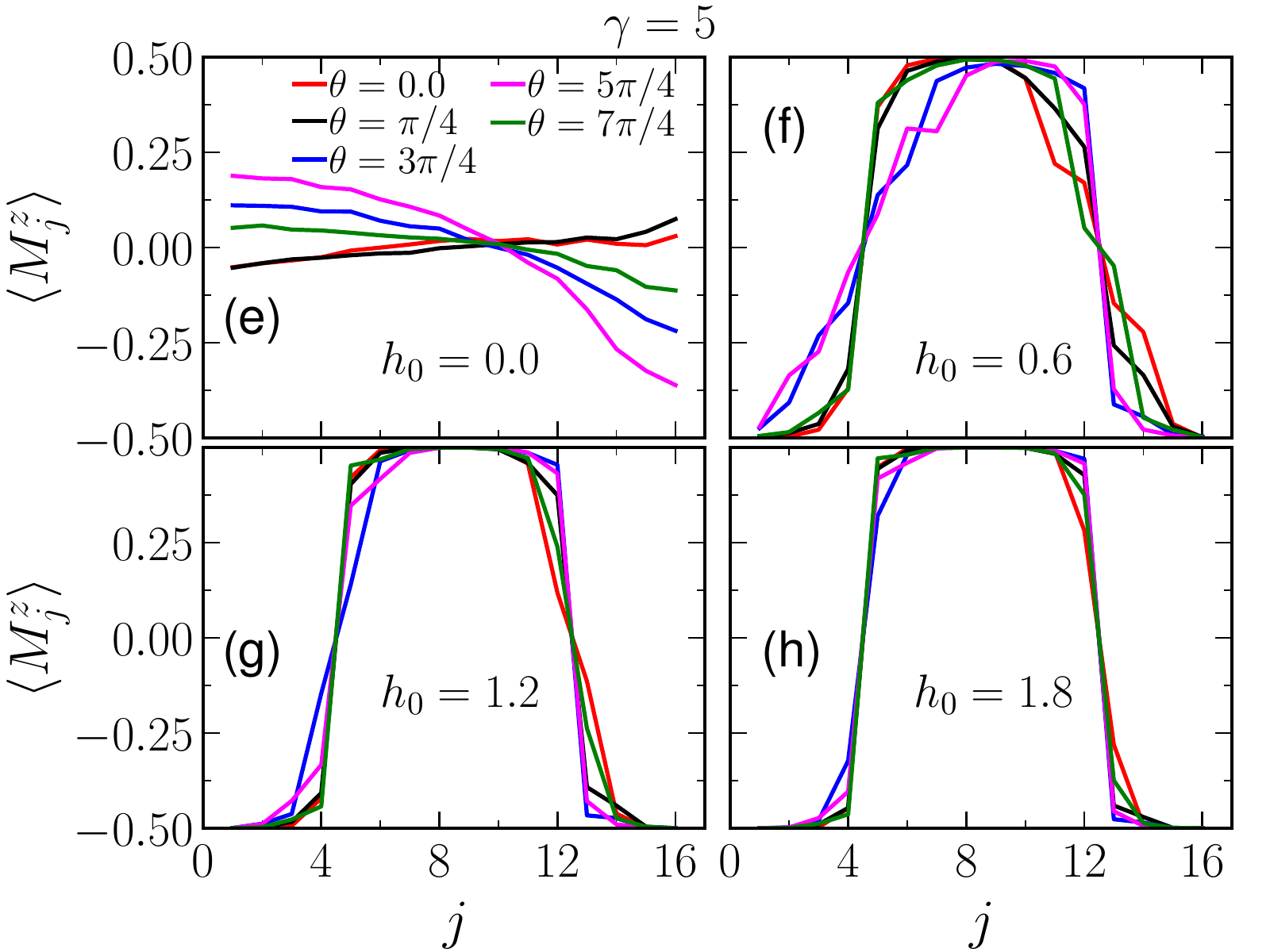}} 
	\caption{ Magnetization $M_j^z$ vs $j$ for various values of $\theta$ and for $h_0=0$ (a), $h_0=0.6$ (b), $h_0=1.2$ (c)  and $h_0=1.8$ (d) using $\gamma=0$. Panels (e)--(h) show  the same  as in (a)--(d) but for $\gamma=5$. For all panels we have used  $N=16$.} 
	\label{fig6}
\end{figure*} 

To see how the domain walls melt at long time, in Fig.~\ref{fig6} we  show the spatial resolved local magnetization $\langle S^z_j\rangle $  along the chain, calculated at large $t$.  As we saw before, even at long times, the physical quantities oscillate about some stationary value. For instance, note the oscillation of the ${\cal I}(t)$ about a value near $0.8$ for $h_0=1.2$ shown in Fig.~\ref{fig4}(c). This oscillation results from small fluctuation of magnetization of sites very close to the domain walls. This can be observed in  the magnetization for different  times $t$ large.  To eliminate these oscillations we average the magnetization  over the $N_{\rm avg}$ values of $t$ taken at later times. More precisely, this time average is calculated as
\begin{eqnarray}\label{t_avg}
M^z_j=\frac{1}{N_{\rm avg}}\sum_{k=1}^{N_{\rm avg}}\langle S^z_j(t_k)\rangle,  
\end{eqnarray}
where we have taken $N_{\rm avg} =50$ and $t_k$ evenly distributed within the interval $[450/J_0,500/J_0]$. This is shown to be good enough to eliminate transient early configurations.

In Figs.~\ref{fig6}(a)--\ref{fig6}(d) we show the resulting magnetization profile $M_j$ for $\gamma=0$ and  various values of $h_0$. Note that for $h_0=0$  [Fig.~\ref{fig5}(a)] $M^z_j$ is almost zero everywhere, showing a full thermalization of the initial state, consistent with the imbalance and entanglement entropy discussed above. For finite $h_0$ we observe that information of the initial state is present at later times. For instance, note that for $h_0=0.6$ [Fig.~\ref{fig5}(b)] the initial magnetization island at the center of the chain is still visible, but the hard domain wall is softened already. As $h_0$ further increases [Figs.~\ref{fig5}(c) and \ref{fig5}(d)] the system becomes more localized, which is manifested as a more robust domain wall. Also, consistent with  we have previously observed for $\gamma=0$, note that depending on the value of $\theta$, the left or right domain wall softens  faster. The rotation symmetry $\theta\rightarrow \theta +\pi$ is also observed as the shape of $M^z_j$ is reflected about the center of the chain. The results of $M^z_j$ for  $\gamma=5$ are shown in Figs.~\ref{fig5}(e)--\ref{fig5}(h). It is clear that $M^z_j$ is no longer zero throughout the chain. Instead, we observe that for $h_0=0$ [Fig.~\ref{fig5}(e)] the magnetic moments tends to accumulate on the edges of the chain. The sign of $M^z_j$ at a given position depends $\theta$ but, in contrast to the case of $\gamma=0$, the rotation symmetry is  clearly no longer observed. As $h_0$ increases, the shape of $M^z_j $ across the chain becomes similar to the case of $\gamma=0$, although always asymmetric. The $\theta$ symmetry broken is expected because for finite $\gamma$ the system lacks inversion symmetry about the center of the chain, this asymmetry was not captured in the LSR statistics discussed in the previous section. 
\begin{figure}[!htbp]
	\centering
	\subfigure{\includegraphics[clip,width=3.4in]{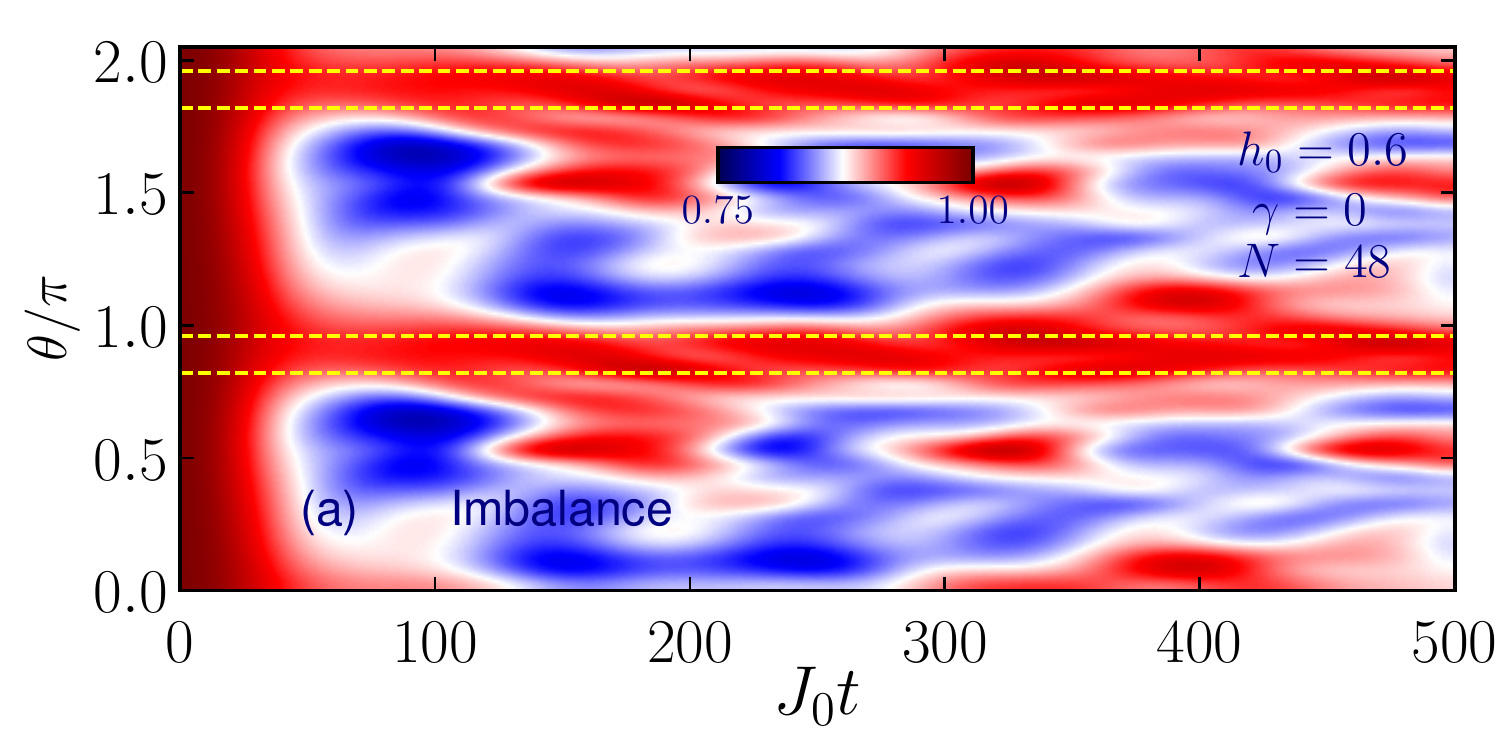}}
\vskip-0.35cm
\subfigure{\includegraphics[clip,width=3.5in]{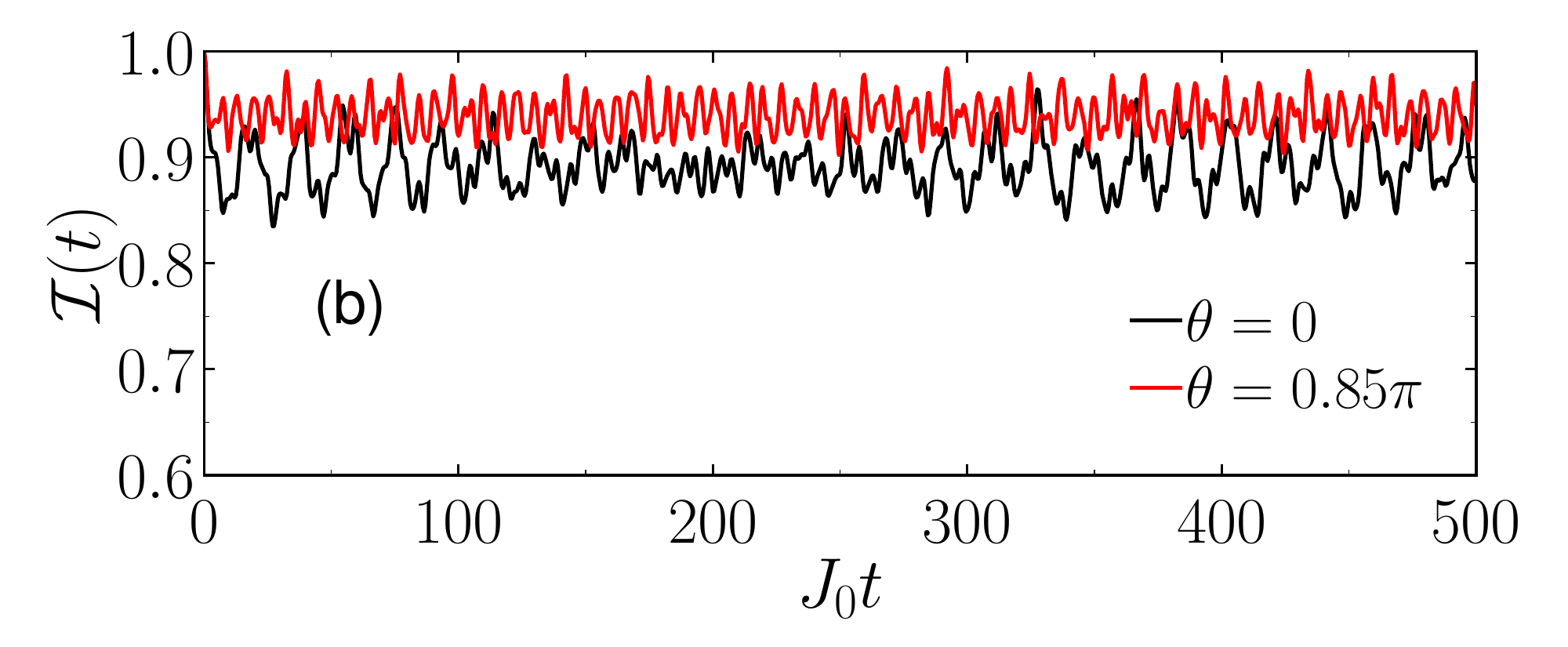}}
\vskip-0.35cm
	\subfigure{\includegraphics[clip,width=3.4in]{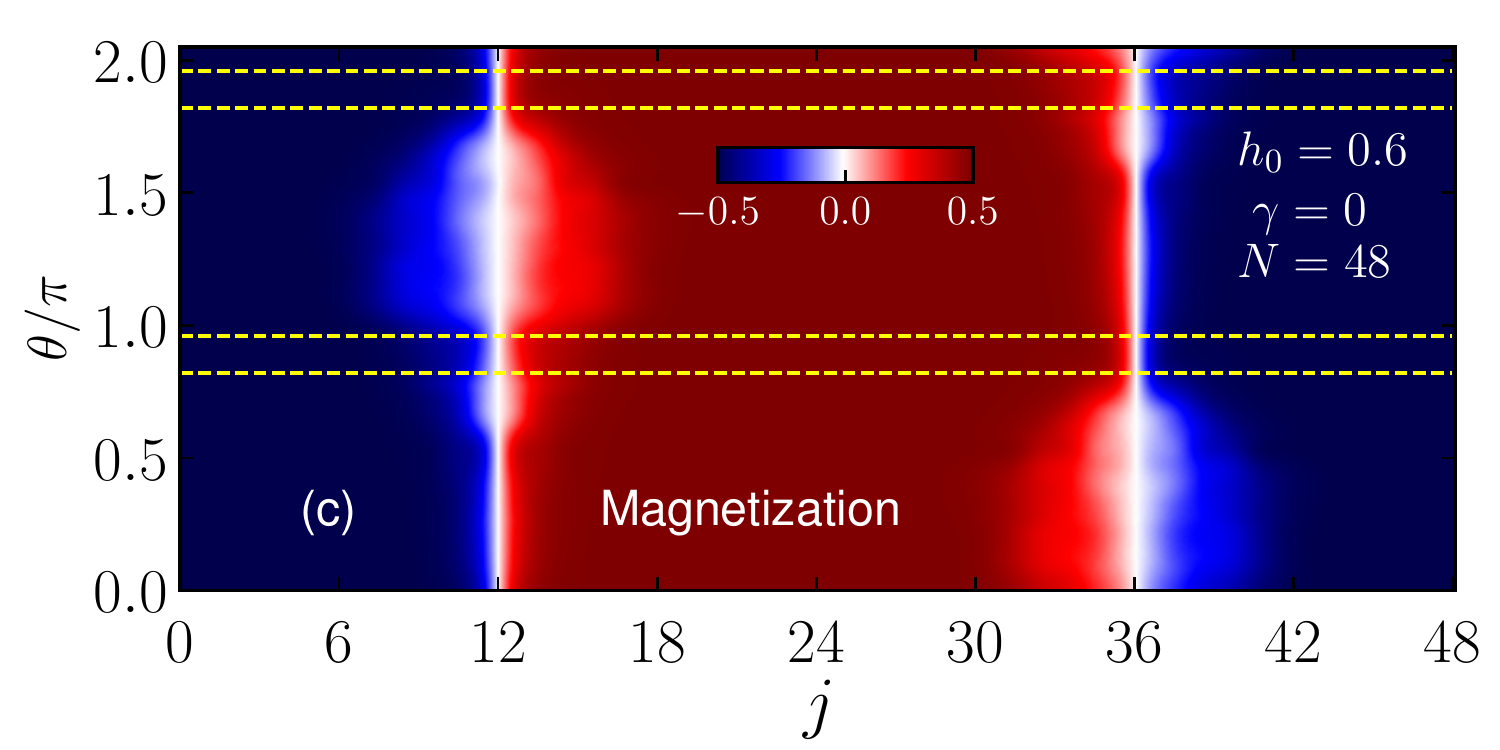}}
	\caption{(a) Color map of the imbalance ${\cal I}(t)$ vs $t$,
	(b) imbalance vs $t$ along a horizontal line of panel (a) for  $\theta=0$ (black) and $\theta=0.85\pi$ (red). (c) Color map of the magnetization $M^z_j$ vs $j$ and $\theta$. In both panels $h_0=0.6$, $\gamma=0$, and $N=48$. Horizontal dashed yellow lines delimit regions in which localization is more robust, as visible as it exhibits sharper left and right domain walls simultaneously.} 
	\label{fig7}
\end{figure} 

Finally, in Fig.~\ref{fig7} we show the results for a chain of length $N=48$ with a field gradient $h_0=0.6$. Figure.~\ref{fig7}(a) shows a color map of the imbalance as a function of $t$ and $\theta$.  We observe the existence  of regions (blue puddles) of small imbalance. This results from some natural frequency in the system --- dependent on $\theta$ --- that induces oscillation in the dynamics of the system. Interestingly, we observe two windows along the $\theta$ axis in which these oscillations are suppressed. To guide the eyes we have delimited these two windows with yellow dashed lines. Within these windows, the time evolution of the imbalance avoids the blue puddles of smaller ${\cal I}(t)$ corresponding to less localized regimes. In Fig.~\ref{fig7}(b) we show ${\cal I}(t)$ vs $t$ for $\theta =0 $ (black) and $\theta=0.85\pi$ (red), corresponding to cuts along horizontal lines drawn in Fig.~\ref{fig7}(a). We note that  ${\cal I}(t)$ is systematically superior for  $\theta=0.85$, which shows an enhanced localization of the state within the region delimited by yellow lines in Fig.~\ref{fig7}(a).

Figure \ref{fig7}(c) shows the color map of the magnetization $M^z_j$ corresponding to the same parameters as in  Fig.~\ref{fig7}(a). We note that all features seen on the left domain wall at a given angle $\theta_0$ is also observed in the right domain wall at an angle $\theta_0+\pi$. In particular, the windows of more robust localization seen in Fig.~\ref{fig7}(a) are also observed here as the softening  of the domain walls is less pronounced in \emph{both} domain walls (this is also delimited with yellow lines). These regions along the $\theta$ axis within which the localization is more robust shows clearly that fine tuning of the next-nearest neighbor can enhance SMBL in the system.

\section{Concluding remarks}\label{conclusions}

In conclusion, we have investigate the robustness of MBL in a one dimensional chain of spin-$1/2$ modeled by a Heisenberg model extended to next-nearest-neighbor isotropic coupling, $J_2$, in addition to a  nonuniform magnetic field. We have employed exact diagonalization and time-evolution calculations that allow to analyze, complementarily, level statistics and thermalization of initial states.
We find that, as discussed in Ref.~\cite{PhysRevLett.122.040606}, for a uniform field gradient, LSR distribution predictions deviates properly describe localized and delocalized regimes of the system. A Similar subtlety has also been reported in Ref.~\cite{PhysRevB.101.134203}.
However, for a slightly nonuniform field gradient, LSR statistics distribution predicts very nicely the delocalized and localized regimes of the system as confirmed by the dynamics of the imbalance and entanglement entropy. Moreover, we show that SMBL is quite robust in the presence of $J_2$ as thermalization is observed for all values of the ratio $J_2/J_1$ (or, equivalently, for all values of $\theta$). More interestingly, we show that there are windows in the $J_2$-$J_2$ plane within which SMBL is enhanced as compared to the special case of $J_2=0$. This feature may be useful to  improve MBL in the context the quantum many-body thermal engine~\cite{PhysRevE.89.032115,PhysRevB.99.024203} as well as in quantum computation~\cite{PhysRevE.62.3504}.  Finally, our results points towards the possibility of more robust SMBL in more general interacting models, including long-range  and noncollinear exchange couplings.


\begin{acknowledgments}
We are thankful to Prof. S. E. Ulloa for very encouraging and enlightening discussions.  We also acknowledge fruitful conversation with Prof. M. Novaes, Prof. F. Pollman, and Z. Bernens. The author also acknowledge support from the Brazilian agencies CNPq (grant N. 305738/2018-6), CAPES and FAPEMIG.

\end{acknowledgments}						


%

\end{document}